\documentclass{aa}
\usepackage{multirow}

\usepackage[T1]{fontenc}
\usepackage{txfonts}
\usepackage{amsmath}
\usepackage{amsfonts}
\usepackage{xcolor}
\usepackage{url}
\usepackage{bm}
\usepackage{graphicx}
\usepackage{amssymb}
\usepackage{soul}
\usepackage{booktabs}
\usepackage{array}
\usepackage[utf8]{inputenc}
\inputencoding{latin1}
\inputencoding{utf8}
\usepackage{appendix}

\def\lsim{ \lower .75ex\hbox{$\sim$} \llap{\raise .27ex \hbox{$<$}} }
\def\gsim{ \lower .75ex \hbox{$\sim$} \llap{\raise .27ex \hbox{$>$}} }

\newcommand{\bi}{\begin{itemize}}
\newcommand{\ei}{\end{itemize}}

\usepackage[colorlinks=true,linkcolor=blue,citecolor=blue]{hyperref}

\begin{document}

\title{The TeV emission of 3C273: inverse Compton radiation from shear-accelerated high-energy electrons in the large-scale jet?}
\titlerunning{Shear acceleration and TeV emission in 3C273}

\author{
Fabrizio Tavecchio\orcid{0000-0003-0256-0995}
}
\authorrunning{Tavecchio}

\institute{
INAF -- Osservatorio Astronomico di Brera, Via E. Bianchi 46, I-23807 Merate, Italy
}
\date{}

\voffset-0.4in



\abstract{The VERITAS Collaboration recently reported the detection of very-high-energy (VHE) gamma-ray emission from the prototypical radio quasar 3C273. The temporal and the spectral properties of this component do not appear compatible with the extrapolation of the beamed blazar-like emission of the inner, pc-scale jet. We explore the possibility that the VHE component is produced in the jet at kpc scale through the inverse Compton emission of a population of ultra-high energy electrons (with Lorentz factor $\gamma\sim 10^8$). In the model these electrons are accelerated through the shear acceleration mechanism, and they also account for the still puzzling X-ray emission of knots detected by {\it Chandra} in the large-scale jets of several powerful quasars (including 3C273). In our scenario the VHE component can be interpreted as the integrated emission from the two brightest knots of the 3C273 jet. We speculate that the decay of the emission on the timescale of  $\sim 3$ years could be accounted for by the scenario if the VHE radiation is produced in some compact regions in the downstream flow of a recollimation shock.}


\maketitle
\boldsymbol{}

\section{Introduction}

Despite two decades of observational and theoretical efforts, the nature of the optical and X-ray emission from compact features (knots) in large-scale jets of powerful radio-loud quasars is still debated (for a review see \citealt{Kraw06}). The widely discussed model which assumes that the X-ray emission is produced by relativistic electrons through the inverse Compton (IC) scattering of the Cosmic Microwave Background (IC/CMB model; \citealt{Tavecchio00,Celotti01}) is strongly challenged by the good upper limits put in the GeV band by {\it Fermi}/LAT (e.g., \citealt{Meyer14,Meyer15}) and by the morphological \citep{Tavecchio03} and  spectral \citep{Jester06} properties of the emission. A potential viable alternative invokes the presence of two distinct electron populations, one at the low energies, responsible for the emission from radio to the optical band, together with an ultra high-energy population (with Lorentz factors $\gamma\sim 10^{7-8}$) emitting X-ray through synchrotron radiation in magnetic fields of $B\sim 10 \; \mu$G (e.g., \citealt{harris02,kataoka05}). This scenario would also explain the high degree of polarization ($>30\%$) measured in the optical band for a component in the jet of the quasar PKS 1136-135, whose optical emission belongs to the high-energy emission component \citep{Cara13}. 

A specific model based on the double-population scenario has been presented in \cite{Tavecchio21}. The model assumes that low energy electrons are accelerated at shocks through the standard diffusive shock acceleration mechanism, while electrons further energized through shear acceleration \citep{rieger19,RiegerDuffy22,Wang23} form a narrow distribution producing synchrotron X-ray emission. The model is able to explain the observed spectral energy distribution (SED) of the knots, with an acceptable energy budget for the jets. 
In the model of \cite{Tavecchio21} the jet is only mildly relativistic, with bulk Lorentz factors $\Gamma\sim 2$, thus avoiding the overproduction of GeV gamma rays through the IC/CMB mechanism as constrained by {\it Fermi}/LAT. However, both the low and high energy electron populations produce  high and very-high gamma-ray radiation through the IC scattering of synchrotron photons and CMB. The resulting non-negligible high-energy emission extends into the TeV band. 

The VERITAS collaboration has recently reported the detection of VHE emission from the prototypical radio-quasar 3C 273\footnote{A presentation can be found at \url{https://indico.cern.ch/event/1258933/contributions/6491204/attachments/3104111/5500883/Benbow_ICRC2025.pdf}}. The light curve shows evidence for a possible variation (i.e. decay) of the flux with a timescale of $\sim$3 years. The ( time-integrated) hard spectrum, raising in the $\nu F(\nu)$ representation, is incompatible with the extrapolation of the soft tail of the blazar $\gamma$-ray component recorded in the GeV band by {\it Fermi}/LAT, associated to the IC emission from the inner, pc-scale jet  (e.g. \citealt{Ghisellini10}). A feasible explanation could be to assume an additional emission component associated to the blazar region (possibly of hadronic origin). In this paper we explore instead the possibility that the VHE component flags the expected integrated IC emission produced by the high-energy, shear-accelerated electrons in the large-scale (kpc) jet.

The paper is organized as follows. In Sect.~\ref{sec:model} we present a sketch of the model, in Sect. 3 we discuss the results and in we conclude in Sect.~\ref{sec:discussion}.

Throughout the paper, the following cosmological parameters are assumed: $H_0=70{\rm\;km\;s}^{-1}{\rm\; Mpc}^{-1}$, $\Omega_{\rm M}=0.3$, $\Omega_{\Lambda}=0.7$.

\section{The model}
\label{sec:model}

We sketch below the main ingredients of the model. For a detailed description the reader is referred to \cite{Tavecchio21}.

As mentioned above, we assume that the low energy (radio to optical) emission of knots in 3C273 is produced by relativistic electrons accelerated by a shock (LE electrons in the following). We assume that these electrons follow a cut-off power-law energy distribution with slope $n_{\rm sh}$: 
\begin{equation}
n_0(\gamma)=K \gamma^{-n_{\rm sh}} \exp\left(-\frac{\gamma}{\gamma_{\rm cut}} \right); \;\;\; \gamma> \gamma_{\rm min}   
\label{eq:n0}
\end{equation}
where $K$ is a normalization and $\gamma_{\rm cut}$ is the maximum Lorentz factor reached by the particles. 

The jet is characterized by a velocity structure, with a core of constant speed surrounded by a shear layer in which the plasma speed $\beta_j(r)$ decreases along the radial direction $r$. The shear could be the result of the interaction of the jet with the external material and the triggering of Kelvin-Helmoltz instability (e.g. \citealt{Borse2021,Wang23}).
We use the following simple linear parametrization for the velocity profile:
\begin{equation}
\beta_j(r)=\left\{ \begin{array}{ll}
                    \beta_{j,0}  &  \mbox{$r \leq r_j-\Delta L$} \\
		    \beta_{j,0} - \frac{\beta_{j,0}}{\Delta L} \left(r-r_j+\Delta L\right)  &  \mbox{$r> r_j-\Delta L$},
		   \end{array} 
		   \right. 
\end{equation}
where $r_j$ is the jet radius and $\Delta L$ characterizes the thickness of the layer. The precise shape of the profile can affect the resulting electron energy distribution, e.g. \cite{RiegerDuffy22}. However, considering the limited quality of the data, the particular choice is not expected to have a major impact on the model.
A small fraction of the LE electrons accelerated at the shock enter the sheath in the downstream region, where they experience further energization through scattering with magnetic turbulence (HE electron component in the following).

For the modeling of the shear acceleration process we follow \cite{Liu17}, which uses an approximate Fokker-Planck treatment to model the time-dependent evolution of the electron energy distribution.
A key parameter regulating the shear acceleration process is the mean free path of particles scattered by magnetic turbulence, $\lambda$, that can be written:
\begin{equation}
\lambda(\gamma)=\frac{r_g}{\xi} \left(\frac{r_g}{\Lambda}\right)^{1-q},
\end{equation}
where $\gamma$ is the particle Lorentz factor, $r_g$ is the particle gyration radius, $\xi=\delta B^2/B^2$ is the ratio between the energy density of the turbulent and the regular magnetic field, $\Lambda$ is the maximum wavelength of turbulence interacting with the particles and $q$ is the slope of the power law spectrum of the turbulent field in the wavenumber space. In the following we fix  $q=5/3$ , corresponding to the Kolmogorov spectrum.

A second important parameter regulating the acceleration efficiency is related to the velocity profile and it is defined as $A=\Gamma_j(r)^2|\partial_r \beta_{j}(r)| c$, where $\Gamma_j(r) =[1-\beta_{j}(r)^2]^{-1/2}$ is the flow bulk Lorentz factor.
The acceleration timescale can be written as: 
\begin{equation}
t_{\rm acc}(\gamma)=\frac{15}{6-q}\frac{c}{\lambda A^2}.
\end{equation}
Note that $t_{\rm acc}\propto (\gamma/B)^{-1/3}$ for $q=5/3$, therefore $t_{\rm acc}$ decreases with the particle energy, i.e. particle at large energies are more efficiently accelerated than low energy ones. This is the reason why a population of pre-accelerated electrons is required for this mechanism to work efficiently. On the other hand, large magnetic fields, determining small gyroradii and thus small $\lambda$, imply large acceleration times. 

Particles can also diffusively escape from the acceleration layer. We approximated the escape time considering the diffusion of particles from a region with size comparable to the shear layer thickness $\Delta L$, i.e., $t_{\rm esc}(\gamma)\simeq \Delta L^2/2\kappa$, 
where for the spatial diffusion coefficient one can use the standard expression $\kappa=c\lambda/3$. 
To ensure acceleration, we have to assume $t_{\rm acc}<t_{\rm esc}$.

At very high-energy the mean free path of the electrons becomes comparable to the size of the jet layer, $\Delta L$. Beyond this point the particles escape from the system and cannot be further accelerated. This fixes a geometrical maximum limit $\gamma_{\rm g,max}$ for the accelerating electrons,  given by $\lambda(\gamma_{\rm g,max})=\Delta L$. 

Both LE and HE electrons emit through synchrotron and IC mechanisms. For the latter we consider both photons locally produced in the jet and CMB photons. The radiative losses determine a second (radiative) limit to the maximum energy reached by the electrons, $\gamma_{\rm r,max}$, set  by the balance between cooling and acceleration timescales. The radiative cooling time is:
\begin{equation}
t_{\rm cool}(\gamma)= \frac{3 m_ec}{4 \sigma_T (U_B+U_{\rm rad})\beta^2\gamma}
\end{equation}
where $U_B$ is the magnetic field energy density and the  total radiation energy density includes the contribution of low-energy synchrotron photons emitted by electrons in the downstream region and the CMB photons, $U_{\rm rad}=U_{\rm syn}+U_{\rm CMB}$ (we neglect the IC emission off the synchrotron photons produced by HE electrons because scatterings occur deeply into the Klein-Nishina regime). 
As we will see, in the case considered here the limiting energy set by the radiative losses is much larger than that determined by the mean free path. The maximum energy of the electrons is thus determined by the geometrical limit, $\gamma_{\rm g,max}$.

The treatment used here adopts a leaky-box, spatially averaged approximation of the shear acceleration process \citep{RiegerDuffy22} that leads to a Fokker-Planck equation prescribing the evolution of the electron energy distribution \citep{Liu17}. As in the case of the velocity profile, the details of the spatial transport are not expected to be highly relevant in the present context. Numerically, we solve the equation by using the robust implicit method of \cite{ChangCooper}.  We assume that a fraction of the particles downstream of the shock, characterized by the distribution $n_0(\gamma)$ defined by Eq. \ref{eq:n0}, enters the shear acceleration process. The injection rate can be phenomenologically described by an injection time $\tau_{\rm inj}$, related to the diffusion time in the downstream region. For simplicity we assume an energy independent timescale.  Therefore we assume a constant injection with spectrum $Q(\gamma)=n_0(\gamma)/\tau_{\rm inj}$. As shown in \cite{Tavecchio21}, to reproduce the hard optical-X ray continuum traced by the data, the lifetime of the source has to be large enough to reach an equilibrium state. For smaller times, the incompletely developed bump displays a relatively soft spectrum incompatible with the optical and X-ray fluxes. Therefore, we calculate the HE component until an approximate equilibrium state is reached.

\begin{figure}[t]
    \centering
    \hspace{0. truecm}
    \vspace{-0.25 truecm}
      \includegraphics[width=0.9\linewidth,height=0.75\linewidth]{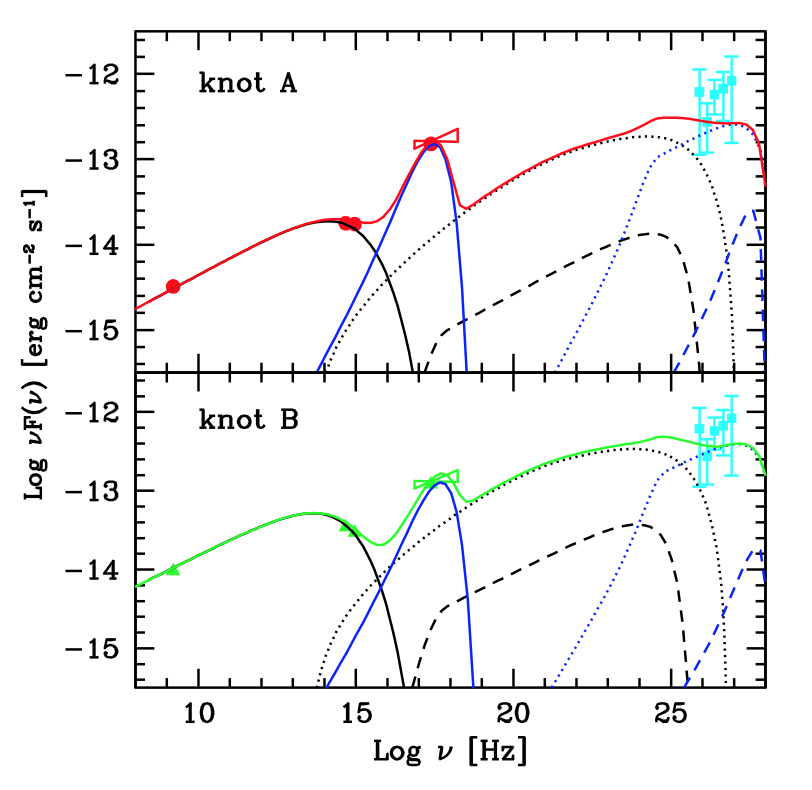}
    \caption{SED (circles) of knot A (upper panel) and B (lower panel) of the jet of 3C273 (from \citealt{Sambruna01}: {\bf we also report the X-ray spectral slopes derived in \citealt{He23}}). Light blue datapoints report the (EBL-deabsorbed) VERITAS spectrum. The lines show the result of the model for the electrons accelerated at the shock (black) and at the shear (blue). Solid: synchrotron; dotted IC from synchrotron; dashed: IC/CMB. The red and the green lines report the total emission.}
    \label{fig:combo}
\end{figure}

For the application to 3C273 we fix some parameters to benchmark values. In particular we assume $B=10$ $\mu$G (both in the core and in the sheath), $\Gamma_j(0)=2$ for both knot A and B. We assume a viewing angle of 15 degrees, implying a Doppler factor (relevant for the boosting of the emission) $\delta=3$.
For definiteness we also fix $\Delta L=0.2 r_j$, $\xi=0.1$ and $\Lambda=\Delta L$. The assumption that the observed decay timescale of the VHE emission, $t\sim 3$ years, is dictated by the light-crossing time of the emitting region (see discussion below), together with the assumed $\delta$, implies a dimension of the source (that in our scenario should be identified with the shear layer) $\approx 10^{\rm 19}$ cm.

\section{Results}
\label{sec:results}

Tuning the free parameters, we first reproduce with the LE component the radio-to-optical SED of the two brightest knots, called knot A and B by \cite{Sambruna01}. The second emission component, produced by HE electrons accelerated within the velocity shear, is matched to the X-ray data by regulating the only remaining free parameter, the injection timescale, $\tau_{\rm inj}$, for which in both cases we derive $\tau_{\rm inj}\sim 10^2$ yrs. 

We show the datapoints (radio, optical, X-rays) in Fig.\ref{fig:combo}, together with the results of the modelling (whose parameters are reported in Tab.\ref{tab:results}). In the plot we show all radiative components from both LE and HE electron populations. In particular, the synchrotron  radiation of the LE electrons accounts for the radio-to-optical components, while the HE population produce the X-ray emission. Both IC components from LE and HE contribute to the emission in the gamma-ray band. Note that the IC scattering of HE electrons with the synchrotron photons produced by themselves is deeply into the Klein-Nishina regime and therefore it is strongly suppressed. HE electrons therefore emit IC radiation mainly through the scattering of the CMB and the synchrotron photons emitted by LE electrons. As visible in Fig.\ref{fig:combo}, the main contribution to the emission in the TeV band comes from the latter component. We remark that the X-ray emission corresponds for both knots to the peak of the HE synchrotron component. This agrees with the results of \cite{Jester06}, who noted that the X-ray emission of knots in 3C273 is softer than the corresponding radio spectrum. 

\begin{figure}[t]
    \hspace{-0.4 truecm}
    \vspace{-0.2 truecm}
      \includegraphics[width=0.9\linewidth]{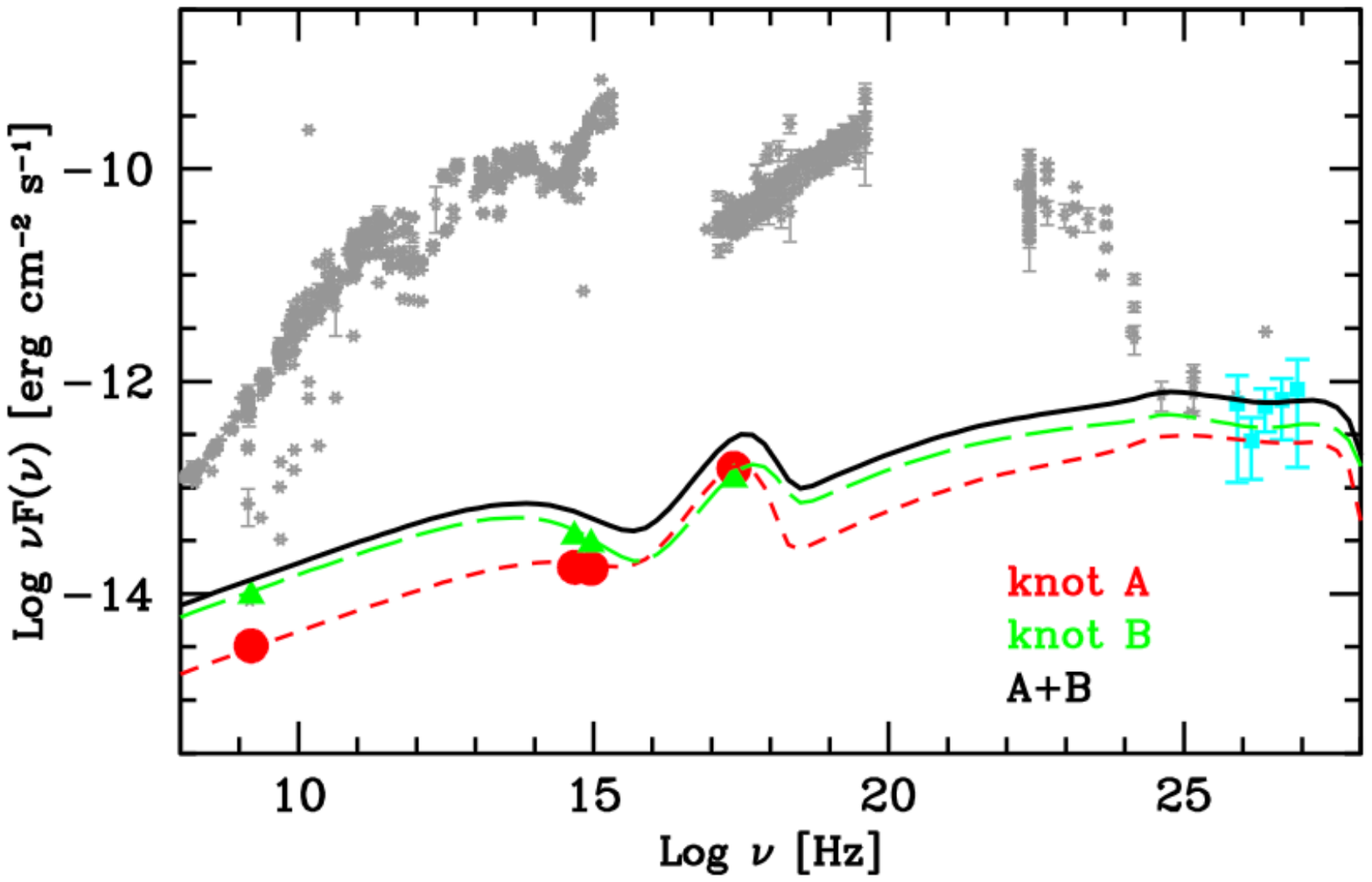}
    \caption{Historical SED data for the 3C273 core (gray) with the VERITAS spectrum (light blue) and the SED of knot A and B. Red, green and black lines show the emission from knot A, B and the total, respectively. }
    \label{fig:tot}
\end{figure}

 In Fig. \ref{fig:tot} we show the SED of the 3C273 core (gray datapoints), together with the VERITAS datapoints. The black curves show the integrated emission from knot A and B, which accounts for the observed VHE component.

\begin{table}[t]
\centering
\begin{tabular}{ccccc}
\hline
\hline
Knot & $\gamma_{\rm cut}$ $(\times 10^6)$ & $K$ & $n_{\rm sh}$ & $r_j$  \\
\quad  & [1] & [2]  & [3] & [4]   \\
\hline
A & 3.5 & $10^3$ & 2.6 & 7   \\
B  & 2 & $1.5\times 10^3$& 2.6 & 9    \\
\hline
\hline
\end{tabular}
\vskip 0.4 true cm
\caption{
Parameters of the models.
[1] cut-off electron Lorentz factor of the shock component;
[2]: normalization of the the shock electron energy distribution (particle cm$^{-3}$);
[3]: slope of the the shock electron energy distribution;
[4]: jet radius ($\times 10^{19}$ cm).
}
\label{tab:results}
\end{table}

For the model we use $r_j<10^{20}$ cm, so that the width of the shear layer, $\Delta L=0.2r_j$ (that in our model quantify the size of the emission region of HE electrons and hence detemines the variability timescale of the VHE component), is close to the constraint posed by variability ($\approx 10^{19}$ cm when beaming is taken into account). 

In Fig. \ref{fig:times} we show the timescales of the different processes as a function of the electron Lorentz factor for knot A. Clearly the radiative cooling time (dominated by the IC losses, dotted blue line) greatly exceeds the acceleration timescale at all energies. Therefore, radiative losses do not limit the acceleration. Instead, the maximum energy of the shear-accelerated electrons is set by the geometrical limit (dashed vertical line) at $\gamma _{\rm g,max}\simeq 10^8$.

With the physical parameters derived from our models the inferred power carried by the jet (including one cold proton per relativistic electron) is $P_J=3\times 10^{46}$ erg s$^{-1}$, smaller than the power derived for the inner jet from the modelling of the SED of the core (e.g. \citealt{Ghisellini10}).

\section{Discussion}
\label{sec:discussion}

We have shown that a scenario based on two populations of relativistic electrons \citep{Tavecchio21} can satisfactorily reproduce the multifrequency emission of the two brightest knots of the kpc-scale jet of 3C273 and can naturally account for the hard very-high energy spectrum measured by VERITAS, interpreted as the integrated IC emission of the knots.
The specific scenario that we adopt assumes that the two electron populations are produced through the joint action of diffusive shock acceleration and shear-acceleration mechanisms. A model based on shear acceleration has been already applied to interpret the emission of large scale jets in radio-loud quasars and radiogalaxies \citep{Tavecchio21,He23,Wang23}. The two latter works also considered the emission from 3C273, but the models predict a flux in the TeV band well below that detected by VERITAS.

\begin{figure}[t]
    \hspace{-0.4 truecm}
    \vspace{-0.2 truecm}
      \includegraphics[width=1.02\linewidth]{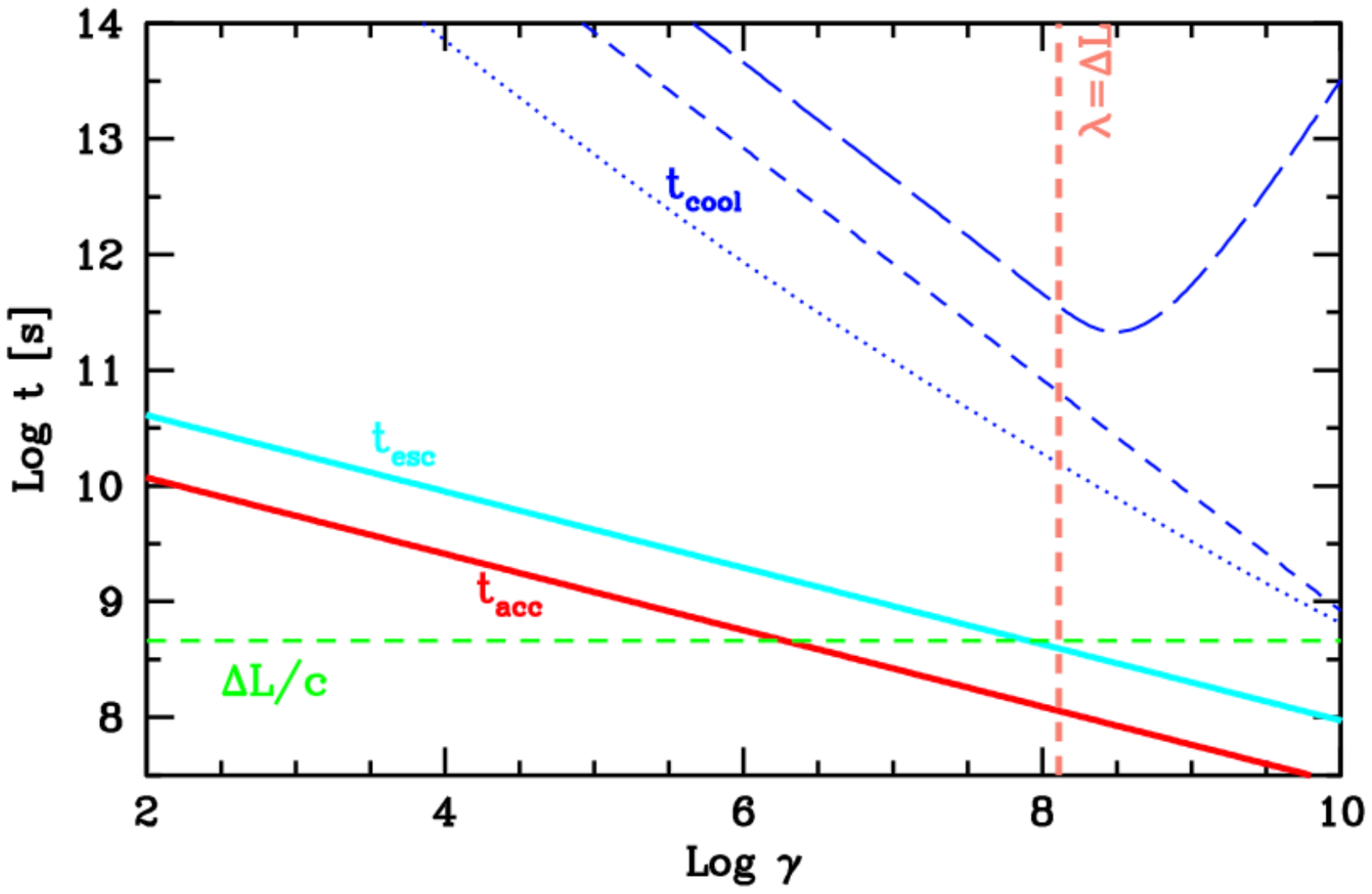}
    \caption{Timescales (in the source frame) relevant for the shear acceleration process as a function of the particle Lorentz factor, for knot A. Solid lines show the acceleration (red) and escape (light blue) timescales. We report (in blue) the curves showing separately the cooling time for synchrotron, IC on the synchrotron photons and IC/CMB (dashed, dotted and long-dashed lines, respectively). The vertical orange dashed line shows the Lorentz factor above which the mean free path exceeds the width of the shear layer, halting the acceleration process. The horizontal dashed green line indicates the light crossing time of the shear layer.}
    \label{fig:times}
\end{figure}


The most challenging constraint to the model comes from the decay timescale of the recorded VHE flux, of the order of about 3 years, i.e. $t_{\rm dec}\sim 10^8$ s. This directly limits the (Doppler corrected) size $R$ of the emitting region $R\lesssim 10^{19}$ cm. In the scheme, the region responsible for the VHE radiation is identified with the shear, with size $\Delta L=0.2 r_j$, thus posing an upper limit $r_j\lesssim10^{20}$ cm to the jet radius. 

Furthermore, to account for the observed decay, we have to assume that particles emitting at VHE, with $\gamma \sim 10^8$, must either cool or escape from the source in a time smaller than $t_{\rm dec}\delta \simeq 3\times 10^8$ s. Although radiative cooling is quite inefficient, Fig. \ref{fig:times} shows that particles can cool through adiabatic losses (of the order of the light crossing time $\Delta L/c$) or diffuse out of the shear layer in a comparable time. 

We note that the compact dimensions of the source,  a constraint primarily determined by the decay timescale  of the observed flux, naturally ensures a high output through the IC component(s), a key ingredient to reach the relatively high flux required by the VERITAS data. We also remark that in our scenario the variability of the VHE emission should reflect in the X-ray band, since HE electrons are responsible for the emission in both bands. To our knowledge, there are no {\it Chandra} observations of 3C273 in the period corresponding to the VHE flux decay, therefore it is not possible to directly test this prediction. However, variability of the X-ray knot emission on year timescale has been detected in other jets, most notably Pictor A \citep{Marshall10}.

The small size of the emission regions flagged by the short-term variability clearly represents a challenge for our understanding of the emission from large scale jets (see also \citealt{Marshall10}). In the specific case of 3C273, knots A and B, the sites of the high-energy emission, lie at distances of the order of $10^{23}$ cm from the central BH (de-projected with the assumed viewing angle of 15 deg). Assuming a simple conical geometry for the jet, this would imply an implausibly small jet opening angle $\theta_j\sim 10^{-3}$ (at VLBI scales, observations suggest $\Gamma_j\theta_j=0.2$, \citealt{Pushkarev09}, that in our case would provide $\theta_j=0.1$). 
As a possible way out from this conundrum, we advance the hypothesis that knots mark the position of strong recollimation shocks (e.g. \citealt{KomissarovFalle97}). These structures form when the imbalance between the jet and the external pressures forces the jet streamlines to bend toward the jet axis, forming an oblique shock followed by a reflection shock. In the complex region downstream of the recollimation shock (where the jet reaches its minimum radius and the reflection shock starts) particles can be accelerated and emit (e.g. \citealt{Sciaccaluga25}). MHD simulations (e.g. \citealt{Matsumoto21,costa24}) show that most of the dissipation (and hence emission) occurs after the reflection shock, in very compact regions, potentially much smaller than the jet radius \citep{BodoTavecchio18}. Moreover, the flow in these regions develops a complex structure, with steep velocity gradients which can potentially provide the ideal conditions for efficient shear acceleration. Dedicated simulations beyond the scope of this paper, including the details of the acceleration processes, should be  performed to assess this scenario.

The upcoming Cherenkov Telescope Array, with its improved sensitivity, will be able to easily confirm the VERITAS detection and to provide high-quality data that can be contrasted with models. Particularly interesting will be the investigation of the temporal behavior of the emission that, as discussed above, puts strong constraints to models.
Among powerful, large scale quasar jets with bright knots, 3C 273 is one of the sources closest to the Earth ($z=0.158$). This, besides implying a large flux, determines a relatively small absorption of the VHE radiation through the pair production process with the EBL \citep{Franceschini17}. Sources located at larger distance (e.g. the prototypical source PKS 0637-752 at $z=0.651$) are likely characterized by a lower flux and are further penalized by a larger EBL suppression. Although dedicated simulations of the observational feasibility are required, one can tentatively conclude that detection of VHE emission from other large-scale jets hosted by powerful quasars is probably very challenging.

\begin{acknowledgements}
I thank the referee for useful comments. I would like to thank S. Boula and P. Coppi for useful and encouraging discussions. FT acknowledges financial support from INAF Theory Grant 2024 (PI F.~Tavecchio) This work has been funded by the European Union-Next Generation EU, PRIN 2022 RFF M4C21.1 (2022C9TNNX). Part of this work is based on archival data provided by the ASI-SSDC.
\end{acknowledgements}

\bibliographystyle{aa}
\bibliography{tavecchio}


\end{document}